\documentclass[onecolumn,amsfonts,amssymb,amsmath,preprint]{revtex4}
\usepackage{SIunits,graphicx}
\begin{document}

\title{Decay dynamics of quantum dots influenced by the local density
  of optical states of two-dimensional photonic crystal membranes}

\author{B. Julsgaard, J. Johansen, S. Stobbe, and T. Stolberg-Rohr}

\affiliation{DTU Fotonik, Department of Photonics Engineering,
  Technical University of Denmark, {\O}rsteds Plads 343, 2800
  Kgs.~Lyngby, Denmark.}

\author{T. S\"unner, M. Kamp, and A. Forchel}

\affiliation{Technische Physik, Universit\"at W\"urzburg, Am
  Hubland, D-97074 W\"urzburg, Germany.}

\author{P. Lodahl}
\affiliation{DTU Fotonik, Department of Photonics Engineering,
  Technical University of Denmark, {\O}rsteds Plads 343, 2800
  Kgs.~Lyngby, Denmark. Website: {\tt
    http://www.fotonik.dtu.dk/QuantumPhotonics}}.

\date{\today }

\begin{abstract}
  We have performed time-resolved spectroscopy on InAs quantum dot
  ensembles in photonic crystal membranes. The influence of the
  photonic crystal is investigated by varying the lattice constant
  systematically. We observe a strong slow down of the quantum dots'
  spontaneous emission rates as the two-dimensional bandgap is tuned
  through their emission frequencies. The measured band edges are in
  full agreement with theoretical predictions. We characterize the
  multi-exponential decay curves by their mean decay time and find
  enhancement of the spontaneous emission at the bandgap edges and
  strong inhibition inside the bandgap in good agreement with local
  density of states calculations.

\end{abstract}

\maketitle

Photonic crystals (PCs) are of fundamental importance due to their
ability to modify the dynamics of light-matter interaction
\cite{Yablonovitch1987a}. The radiative lifetime of internal light
sources can be controlled and the spatial emission pattern can be
modified \cite{Lodahl2004a, Fujita2005a, Kaniber2007a,
  Nikolaev2005a}. These effects open up the possibility of improving
the efficiency and compactness of devices such as light-emitting
diodes (LEDs) \cite{Vahala2003a}, and offer control of the propagation
of single photons of use in, e.g., quantum information protocols
\cite{Simon2007a}. In order to determine the ultimate potential of PCs
in such applications, a thorough understanding of light emitters in
PCs is needed. Here we study systematically the spontaneous emission
dynamics of quantum dots (QDs) in two-dimensional (2D) photonic
crystal membranes (PCMs) and compare our results to theory. The
radiative decay rate of a single QD inside the PCM is governed by the
projected local density of optical states (LDOS) \cite{Sprik1996a}. In
order to probe variations herein, we investigate ensembles of QDs
since in this case any dependence on the property of individual QDs is
averaged out. Furthermore, QD ensembles provide good signal-to-noise
ratio in the measurements and are important for large-volume
applications such as PC lasers, LEDs, or solar cells.  The projected
LDOS varies with the position and orientation of the emitter inside
the PCM \cite{Koenderink2006a}, and a distribution of decay rates is
expected to arise from the QD ensemble with a multi-exponential decay
curve as a result. Such an effect has already been reported in 3D
inverse opals using colloidal QDs \cite{Nikolaev2007a}, but so far a
systematic study of the decay curves in 2D PCMs exists only
theoretically \cite{Koenderink2006a}.

\begin{figure}[t]
  \centering
  \includegraphics[width=\columnwidth]{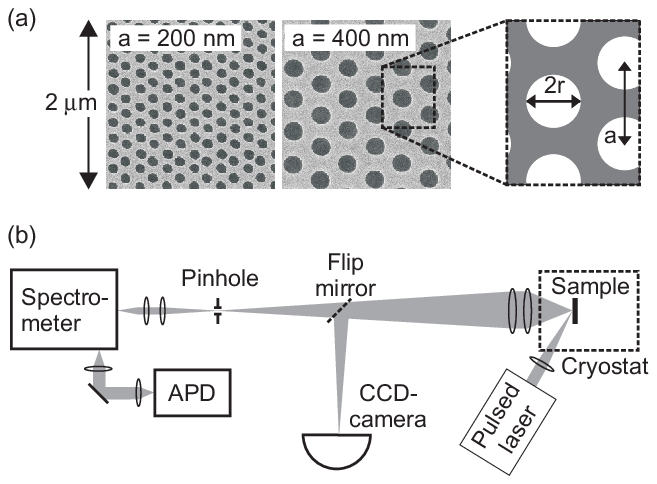}
  \caption{ {\bf (a)} SEM images of the PCMs. {\bf (b)} Experimental
    setup: The PCMs are placed in a cryostat and excited by a pulsed
    laser source. The time-resolved spontaneous emission is detected
    by an avalanche photo diode (APD) after spectral selection. A
    pinhole selects the emission from the central part of the PCM, the
    position of which can be adjusted with assistance from a CCD
    camera.}
  \label{fig:exp_setup}
\end{figure}

We consider ensembles of InAs QDs in GaAs PCMs that have been
fabricated with systematically varied lattice constants. We
optically excite the QDs within the membranes and detect the
time-resolved spontaneous emission from a spectrally selected
ensemble of approximately $6\times 10^3$ QDs.
%
%
%
%
The dimension of each membrane is $40\times 40\:\micro\mathrm{m}^2$
with thickness 150 nm. The QDs are grown by molecular beam epitaxy
with density $\approx 300\:\micro\mathrm{m}^{-2}$ and centered within
the membrane. The PCM consists of holes arranged in a triangular
lattice, see Fig.~\ref{fig:exp_setup}(a), with hole radius, $r$, and
lattice constant (hole spacing), $a$. All membranes in this experiment
are in the range $r/a = 0.313 \pm 0.006$ measured by scanning electron
microscopy (SEM), and the lattice constant ranges from 180 to 470 nm
in steps of 10 nm. The samples are placed in a closed-cycle cryostat
keeping the temperature at 14 K. The QDs in the PCMs are excited in
the GaAs barrier with a PicoQuant PDL-800 pulsed diode laser running
at 781 nm with a repetition period of 100 ns. The chosen pump
intensity corresponds to an upper estimate of 5-10 electron-hole pairs
per QD generated in each excitation pulse, and emission from the
ground state, the bi-excitonic state, and the higher excited states of
the QDs are thus expected. Spectral selection of the ground state
energy ($\lambda = 980\:\mathrm{nm}$) ensures that we only observe
spontaneous emission from the ground and bi-excitonic
states. Saturation of the QDs leads to a 50\% increase of the decay
rate, but this effect is expected to be independent of the PCM
geometry and therefore does not pose a problem to our
measurements. The high pump power improves the signal-to-noise ratio
compared to the case of non-saturated QDs. The experimental setup is
shown in Fig.~\ref{fig:exp_setup}(b).

\begin{figure}[t]
  \centering
  \includegraphics[width=\columnwidth]{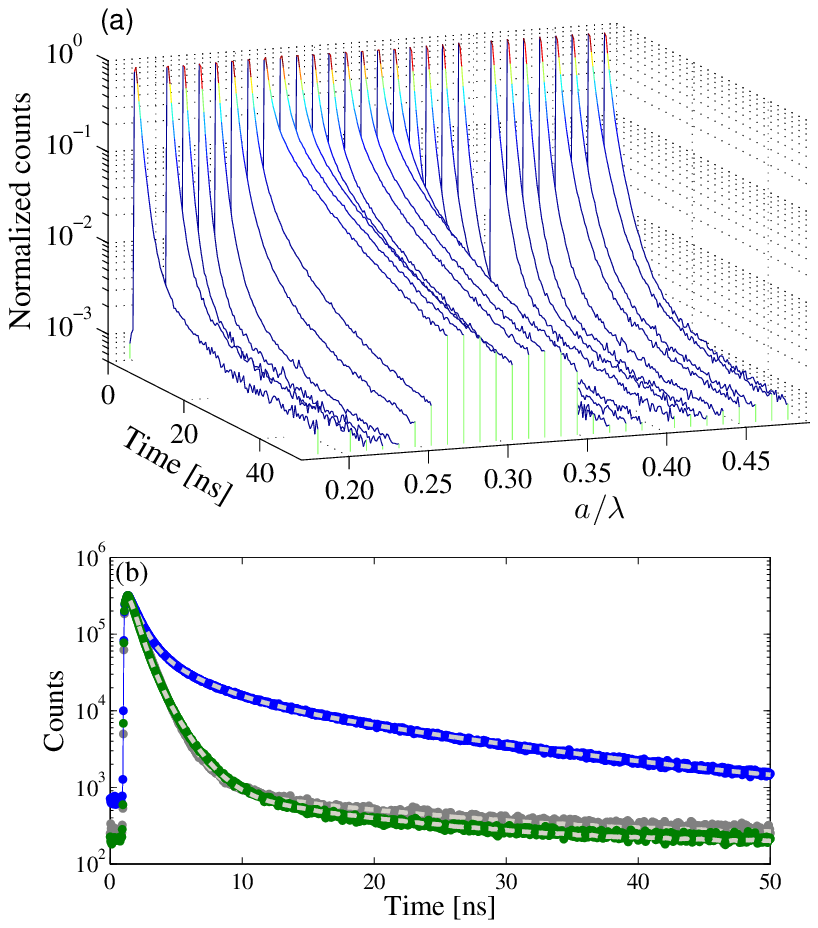}
  \caption{ (color online) {\bf (a)} Time-resolved decay curves
    shown on a normalized scale for varying lattice constant. For
    clarity, only the first 50 ns are displayed.  A 2D bandgap effect
    is clearly visible in the region $0.26 \le a/\lambda \le
    0.35$. {\bf (b)} Fit of the triple-exponential model $f(t)$
    (dashed lines) to the measured decay curves for $a/\lambda=0.296$
    (upper, blue curve) and $0.388$ (lower, green curve). For
    comparison the decay curve of an unpatterned sample is also shown
    (gray curve slightly above the lower curve). All curves are scaled
    to have coincident maxima.}
  \label{fig:waterfall}
\end{figure}

The recorded decay curves are shown on a normalized scale for
comparison in Fig.~\ref{fig:waterfall}(a) with varying lattice
constant, $a$, and fixed detection wavelength, $\lambda =
980\:\mathrm{nm}$, within a bandwidth of 2 nm. We readily observe a
pronounced slow down of the decay dynamics in the range $0.26 \le
a/\lambda \le 0.35$. This slow down is the signature of strongly
inhibited spontaneous emission for emission energies within the 2D
photonic bandgap.  Note that the decay curves span up to three decades
due to the long repetition period of the laser. This allows for a
detailed study of the slow components of the decay curves where strong
inhibition of spontaneous emission will appear.

%
%
As expected from the strong position and orientation dependence of
the LDOS \cite{Koenderink2006a}, the spontaneous-emission decay
curves are multi-exponential. Approximating the initial fast decay
by a single-exponential decay model will only determine the
fastest decay rates in the multi-exponential decay. Hence, a
single exponential model will clearly not suffice in
characterizing the decay dynamics. A suitable measure to quantify
the overall change in emission rate is the inverse of the mean
decay time, $\tau_{\mathrm{m}}^{-1}$, which is computed as:
$\tau_{\mathrm{m}}^{-1} = \frac{\int_0^{\infty}f(t)dt}
{\int_0^{\infty}t\cdot f(t)dt}$, where $f(t)$ describes the
ensemble luminescence. To obtain $f(t)$ we fit a
triple-exponential model to the data. The model has been convolved
with the instrument response function \cite{Lakowicz2006a}
obtained by detecting the pump laser scattered from the sample.
Examples of triple-exponential models fitted to the data are shown
in Fig.~\ref{fig:waterfall}(b). We stress that the individual
parameters are of no physical significance, the triple-exponential
model is merely chosen as it is able to model the luminescence
decay well and thus ensures a reliable extraction of
$\tau_{\mathrm{m}}^{-1}$.

\begin{figure}[t]
  \centering
  \includegraphics[width=\columnwidth]{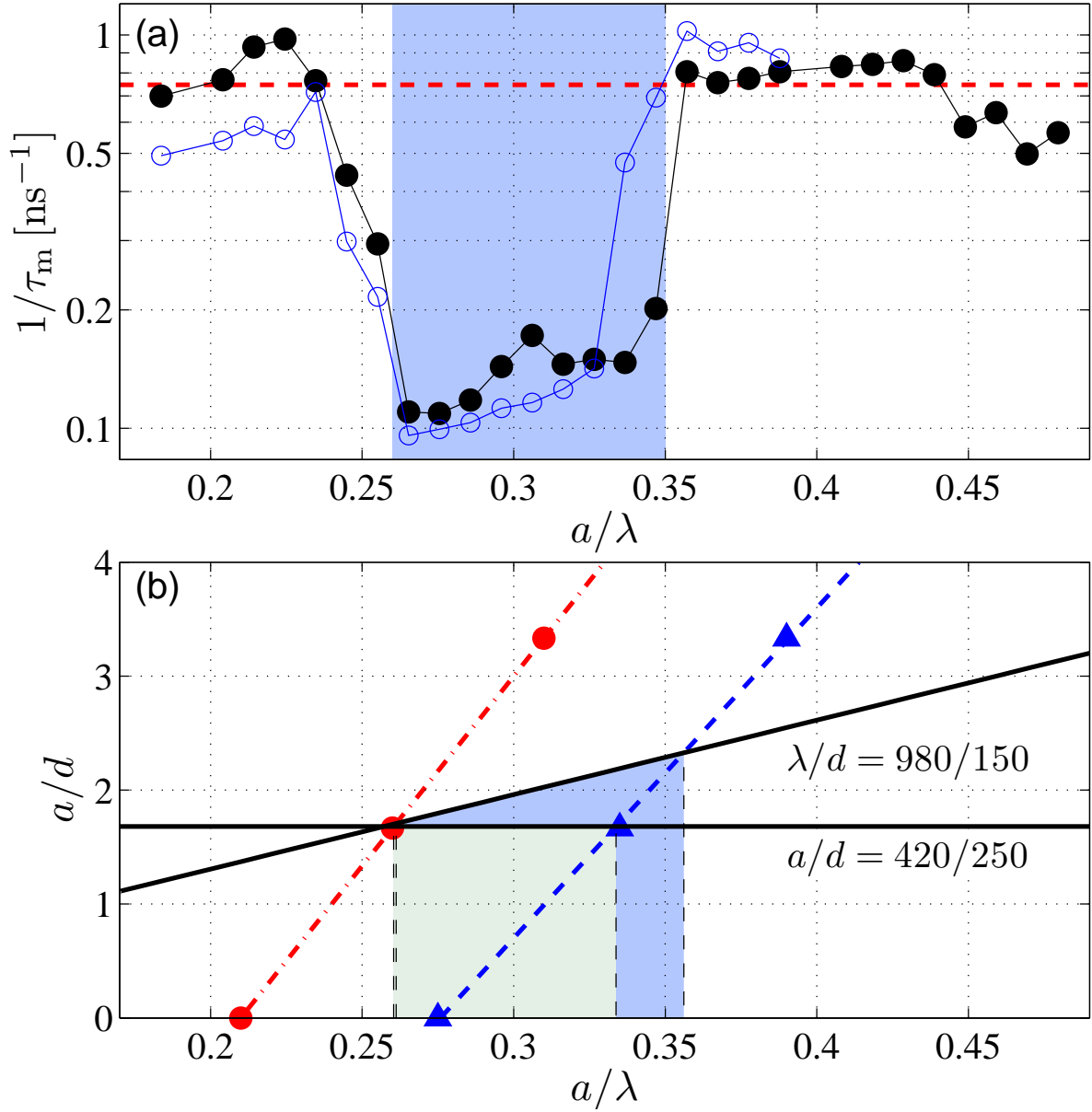}
  \caption{ (color online) {\bf (a)} The inverse mean decay time,
    $\tau_{\mathrm{m}}^{-1}$, of the experimental data (solid, black
    dots) and the simulations (open, blue circles). The dashed, red
    line marks the value of $\tau_{\mathrm{m}}^{-1}$ measured in
    absence of the PCM. The shaded region indicates the measured 2D
    bandgap. {\bf (b)} The position of band edges for various values
    of $a/\lambda$ and $a/d$. Red circles and blue triangles are taken
    from Ref.~\cite{Andreani.JQuantElec.38.891(2002)} with $r/a = 0.3$
    being consistent with the calculations of Koenderink {\it et al.}
    \cite{Koenderink2006a}, which are constrained to the horizontal
    line. The calculated 2D bandgap is indicated by the bright
    shading. Our experiment follows the inclined line with a broader
    2D photonic bandgap as a result (dark shading).}
  \label{fig:MeanRates}
\end{figure}

The inverse mean decay times are plotted in
Fig.~\ref{fig:MeanRates}(a) for the various values of $a/\lambda$.
Compared to the value of $0.75\:\mathrm{ns}^{-1}$ obtained in
absence of the PCM, we observe more than six-fold decrease of the
inverse mean decay time inside the 2D photonic bandgap and an
increase of up to $30\%$ outside the bandgap. It should be noted,
that the effect of variations in the LDOS will be less pronounced
in the mean decay time than in the purely radiative decay time,
since the former also includes contributions from non-radiative
decay, which is unaffected by the LDOS \cite{Johansen2008a}. The
experimental variations in the mean decay times are thus expected
to be a conservative estimate of the actual LDOS variations in the
PCM.

In order to compare the measured mean decay times with theory the
complex QD decay dynamics in a photonic crystal must be adequately
accounted for. The decay curves for an ensemble of QDs will depend on
the position and orientation of the emitter through the LDOS,
redistribution of the emitted light, as well as internal QD
dynamics. Simulation of the data is a comprehensive task and the
details will be described elsewhere. Here we briefly discuss the
approach and show the resulting comparison
(Fig.~\ref{fig:MeanRates}(a)). The radiative decay rate is modeled
using the full 3D LDOS that was calculated for seven different
positions and two orthogonal dipole orientations in
Ref.~\cite{Koenderink2006a}. Redistribution is accounted for by
assuming that modes propagating in the slab are completely inhibited
by the 2D photonic bandgap while radiation modes are unaltered. This
simplistic model has been implemented successfully in the
literature~\cite{Kaniber2007a, Fujita2005a}, however a rigorous
treatment would require calculating the full 3D Green's function of
the PCM, which is not available in the literature.  Finally, internal
QD dynamics is modeled by including the interplay between bright and
dark excitons states, which leads to a bi-exponential decay of the
spontaneous emission \cite{Labeau2003a,Smith2005a}. The relevant
parameters describing the QD dynamics, including non-radiative decay
\cite{Johansen2008a} (which is assumed identical for all QDs and hence
neglecting possible surface recombination effects at the membrane
holes), were derived from measurements performed on an unpatterned
membrane.

As shown in Fig.~\ref{fig:MeanRates}(a) we find a remarkably good
agreement between experimental observations and simulated inverse
mean decay times. This is to our knowledge the first detailed
comparison between experiment and theory for QDs in photonic
crystals. Note in Fig.~\ref{fig:MeanRates}(a) that the measured
spectral position of the upper bandgap edge deviates from the
calculated value based on Ref.~\cite{Koenderink2006a}. However,
this is readily explained by noting that the scaled frequencies of
the band edges, $a/\lambda$, vary linearly with $a/d$ as
calculated in Ref.~\cite{Andreani.JQuantElec.38.891(2002)} where
the spectral positions of the 2D photonic bandgap edges were
determined for three different ratios between the lattice spacing,
$a$, and membrane thickness, $d$. This allows for interpolation to
any value of this ratio, as shown in Fig.~\ref{fig:MeanRates}(b).
The LDOS calculations of Ref.~\cite{Koenderink2006a} were
performed for a constant membrane thickness and lattice spacing,
$a/d = \frac{420\:\mathrm{nm}} {250\:\mathrm{nm}}$, while varying
the scaled emission frequency, $a/\lambda$, corresponding to the
horizontal line in Fig.~\ref{fig:MeanRates}(b). It can be shown
that identical results will be obtained by scaling both $a$ and
$d$ while keeping $\lambda$ fixed. Since it is very inconvenient
to vary the membrane thickness, $d$, our experiment is performed
for fixed detection wavelength, $\lambda$, and membrane thickness,
$d$, while the lattice constant, $a$, is varied. In
Fig.~\ref{fig:MeanRates}(b) this corresponds to the inclined line,
$a/d = a/\lambda\cdot\lambda/d = a/\lambda\cdot
\frac{980\:\mathrm{nm}}{150\:\mathrm{nm}}$, which intersects the
red (dash-dotted) and blue (dashed) band-edge lines very closely
(within experimental precision) to our observed values of 0.26 and
0.35, respectively. While very good agreement between experiment
and theory is obtained, when comparing the inverse mean decay
time, $\tau_{\mathrm{m}}^{-1}$, as is apparent from
Fig.~\ref{fig:MeanRates}(a), we note that increased complexity is
found when comparing the full decay curves to theory (not shown).
We believe that the discrepancies found are primarily caused by
the rather simplistic model for the redistribution of light, while
deviations due to the slight differences in $d$ between experiment
and theory are expected to be of minor importance.

In conclusion, we have carried out a systematic study of the
spontaneous emission dynamics in 2D PCMs. Very pronounced
inhibition of spontaneous emission was demonstrated within the
range of lattice constants for which a 2D photonic bandgap is
predicted by theory. Comparing to a reference measurement on QDs
in unpatterned membranes we report an inverse mean decay time
which is reduced by more than a factor of 6 for emission energies
inside the bandgap, while an increase of $30\%$ was observed on
the red side of the bandgap. Our experiment was compared to theory
taking into account the full 3D LDOS and very good agreement was
observed.

We thank Femius Koenderink for generously sharing his LDOS
calculations and J\o rn M. Hvam for fruitful discussion and general
support. We gratefully acknowledge financial support from the Danish
Research Council (FNU grant 272-05-0083 and 272-06-0138 and FTP grant
274-07-0459). The work is part of the EU project ``QPhoton''.
B.~Julsgaard is supported by the Carlsberg Foundation.

\newpage



\newpage

%

\end{document}